\documentclass[showpacs,aps,prl,twocolumn,superscriptaddress]{revtex4-1}
\usepackage{graphicx} 
\usepackage{dcolumn}
\usepackage{bm}
\usepackage{bbold}
\usepackage{dsfont}
\usepackage{amssymb,amsmath}
\usepackage{epstopdf}
\usepackage{color}
\usepackage{ulem}

\begin{document}
\title{Cyclotron resonance of single valley Dirac fermions in gapless HgTe quantum well}

\author{J.~Ludwig}
\affiliation{National High Magnetic Field Laboratory, Tallahassee, FL 32310, USA}
\affiliation{Florida State University, Tallahassee, Florida 32306, USA}
\author{Yu. B.~Vasilyev}
\email{yu.vasilyev@mail.ioffe.ru}
\affiliation{Ioffe Physical Technical Institute RAS, St. Petersburg, 194021, Russia}
\author{N. N.~Mikhailov}
\affiliation{A.V. Rzhanov Institute of Semiconductor Physics SB RAS, Novosibirsk, 630090 Russia}
\author{J. M.~Poumirol}
\affiliation{National High Magnetic Field Laboratory, Tallahassee, FL 32310, USA}
\author{Z.~Jiang}
\affiliation{School of Physics, Georgia Institute of Technology, Atlanta, GA 30332, USA}
\author{O.~Vafek}
\affiliation{National High Magnetic Field Laboratory, Tallahassee, FL 32310, USA}
\affiliation{Florida State University, Tallahassee, Florida 32306, USA}
\author{D.~Smirnov}
\email{smirnov@magnet.fsu.edu}
\affiliation{National High Magnetic Field Laboratory, Tallahassee, FL 32310, USA}

\date{\today}

\begin{abstract}
We report on Landau level spectroscopy studies of two HgTe quantum wells (QWs) near or at the critical well thickness, where the band gap vanishes. In magnetic fields up to $B$=16~T, oriented perpendicular to the QW plane, we observe a $\sqrt{B}$ dependence for the energy of the dominant cyclotron resonance (CR) transition characteristic of two-dimensional Dirac fermions. The dominant CR line exhibits either a single or double absorption lineshape for the gapless or gapped QW. Using an effective Dirac model,
we deduce the band velocity of single valley Dirac fermions in gapless HgTe quantum wells, $v_F=6.4 \times10^5$~m/s, and interpret the double absorption of the gapped QW as resulting from the addition of a small relativistic mass.
\end{abstract}


\maketitle

Since the Dirac-like character of the low-energy band structure in graphene was first demonstrated \cite{Novoselov:2005, Zhang:2005}, there has been a continuous search for other Dirac-fermion materials. It has been shown that two-dimensional (2D) Dirac fermions can be realized in HgTe-based semiconductor quantum wells (QWs) with an inverted band structure and vanishing band gap \cite{Bernevig:2006, Konig:2007, Konig:2008, Buttner:2011}. Unlike graphene, which has two spin-degenerate massless Dirac cones at two inequivalent corners of the first Brillouin zone, gapless HgTe QWs possess a single spin-degenerate Dirac valley at the zone center, thus offering a model system for studying 2D Dirac fermions without inter-valley scattering \cite{Buttner:2011, Tkachov:2011}. Recent magneto-transport measurements performed on HgTe QWs with a critical well thickness ($d_c$) where the band gap vanishes have provided experimental evidence of a single Dirac valley in the system, and Hall measurements show the anomalous sequence of quantum Hall plateaus specific to 2D Dirac fermions \cite{Buttner:2011}.

In a system of 2D Dirac fermions subjected to a perpendicular magnetic field $B_\perp$, a zero-field continuum of electronic states with the characteristic linear dispersion $E = \pm v_F \hbar k$ transform into a set of unequally spaced Landau levels (LLs) \cite{McClure:1956, Semenoff:1984, Haldane:1988} with energies $E_n =\mathrm{sgn}(n)\sqrt{2|n|}v_F\hbar/\ell_B$, where the band velocity, $v_F$, is the slope of the Dirac cone at $B=0$, and $\ell_B=\sqrt{\hbar c/eB_\perp}$ is the magnetic length. The LL index $n=...,-2, -1, 0, 1, 2, ...$ describes the states in the conduction ($n$$>$0) and valence ($n$$<$0) bands, as well as the unusual $n=0$ LL which is located at zero energy. In graphene, the distinctive $\sqrt{B}$ dependence of LL energies has been probed by infrared (IR) cyclotron resonance (CR) experiments, enabling direct and accurate measurements of the band velocity $v_F$ \cite{Jiang:2007CR, Deacon:2007CR}.

\begin{figure}[t]
\includegraphics[width=8cm]{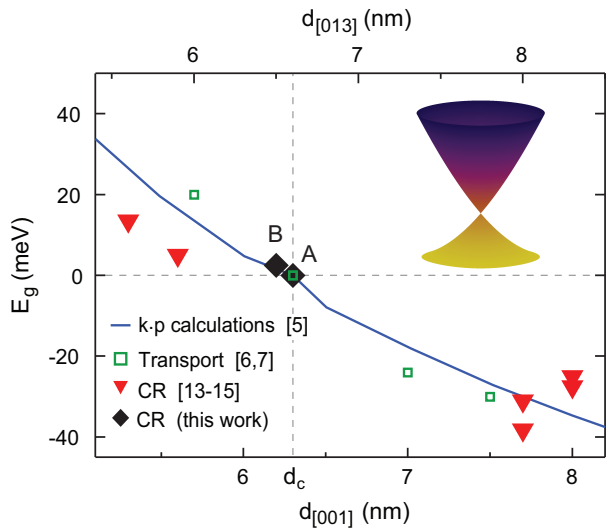}
\caption{
Dependence of the HgTe band gap, $E_g$, on the QW width near $d_c$. The solid line shows $E_g(d)$ from \textbf{k$\cdot$p} calculations in Ref.~[\onlinecite{Konig:2007}] for QWs grown on the [001]-oriented GaAs substrates, yielding $d_c \approx 6.3$~nm. Symbols are extracted from transport (open) or CR (solid) experiments. The bottom axis is for QWs grown on either lattice matched CdZnTe \cite{Tkachov:2011, Buttner:2011} or [001]-oriented GaAs substrates \cite{Konig:2007, Orlita:2011}, while the top axis is for [013]-oriented GaAs substrates \cite{Zholudev:2012, Ikonnikov:2011}.
To compare our results to the $E_g(d)$ dependence for QWs grown on differently oriented substrates, the top axis is shifted by $-0.3$~nm with respect to the bottom axis. Inset: Low-energy band structure of HgTe QW at the critical well thickness.
}
\label{Eg}
\end{figure}

Previous magneto-spectroscopy studies of HgTe QWs have either been performed on samples relatively far from the critical well thickness \cite{Zholudev:2012, Orlita:2011}, $d_c\sim6.2-6.6$~nm, or only probed the inter-LL transitions at low magnetic fields with monochromatic THz light sources \cite{Kvon:2011, Olbrich:2013, Ikonnikov:2011}.
The extracted gap value as a function of nominal QW thickness for both optical and transport studies on HgTe QWs near $d_c$ is summarized in Fig. \ref{Eg}.
In this Letter, we present the low-temperature IR magneto-spectroscopy data obtained from two HgTe QWs at and near the critical thickness corresponding to either a gapless or small-gap state (Fig.  \ref{Eg}).
We observe that the dominant CR lines follow a $\sqrt{B}$ dependence over the whole range of studied magnetic fields, up to 16~T. For both samples, the same slope is extracted from the CR energy versus $\sqrt{B}$ curve, while a single- or double-Lorentzian CR absorptions are observed for the gapless or gapped HgTe QW, respectively. The results are described as CR transitions between the $n$=0 and $n$=1 LLs of 2D Dirac fermions with the same band velocity $v_F=6.4 \times10^5$~m/s and either zero or small, finite relativistic masses.


\begin{figure}[t]
\includegraphics[width=8cm]{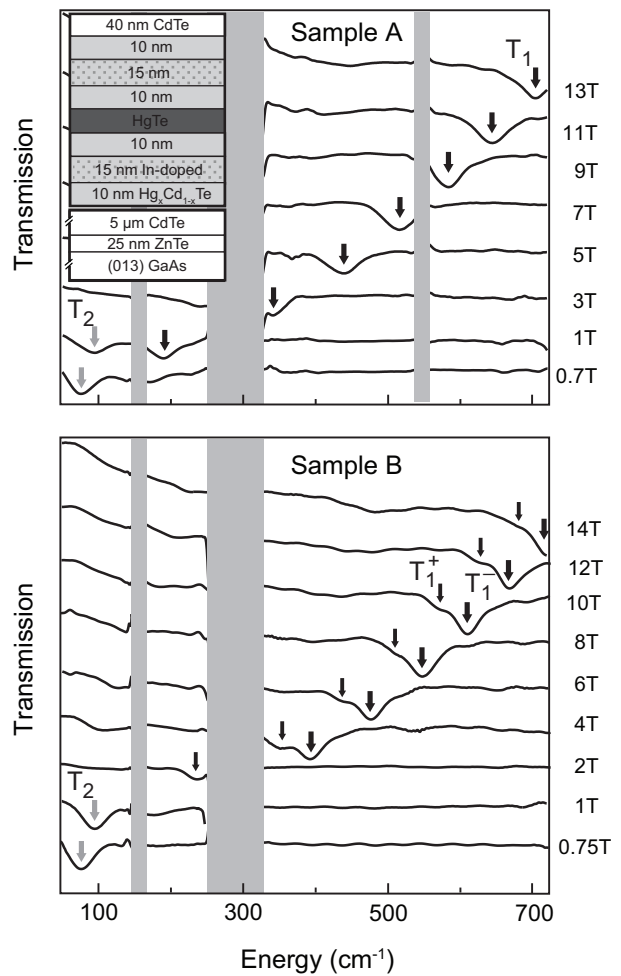}
\caption{
Normalized transmission spectra of HgTe QWs, $T_{B}/T_{0}$, measured at selected magnetic fields. CR absorptions are marked by arrows and spectra are offset vertically for clarity. Inset: Schematic of the structure for the samples measured. The dark gray layer represents the HgTe QW and the light gray regions are the Hg$_x$Cd$_{1-x}$Te barriers.}
\label{FigSpectra}
\end{figure}

The two QW samples studied in this work were grown by molecular beam epitaxy on [013]-oriented GaAs substrates, with nominal well widths of $d=6.6$~nm (sample A) and 6.5~nm (sample B). In order to accommodate for lattice mismatch, buffer layers of ZnTe and CdTe were grown on top of the substrate. A schematic of the QW structure is shown in the inset of Fig. \ref{FigSpectra}. The HgTe QW is sandwiched between Hg$_x$Cd$_{1-x}$Te barriers, $x =0.42$ for sample A and $0.36$ for sample B, and capped with $40$~nm of CdTe. In both samples, the QW is remotely doped on each side by a 15 nm thick In-doped region embedded in Hg$_x$Cd$_{1-x}$Te barriers. The In doping concentration is $2.5$ and $2.7\times10^{16}$~cm$^{-3}$ in sample A and B, respectively. Mobilities greater than $10^{5}$~cm$^{2}$/Vs have been reported in similar remotely doped samples with QWs near the critical thickness and carrier concentrations of about $2\times10^{10}$~cm$^{-2}$ \cite{Kvon:2011, Olbrich:2013}.


IR transmission measurements are performed in the Faraday configuration at 4.2 K with the magnetic field applied perpendicular to the QW plane. The spectra are recorded by a Fourier-transform spectrometer with a resolution of either 4~cm$^{-1}$ or 6~cm$^{-1}$ using a magneto-optical setup similar to that described in Ref.~[\onlinecite{Jiang:2007CR}]. The transmitted IR radiation is measured by a composite Si bolometer mounted directly beneath the sample. The spectra acquired at constant $B_\perp$ are normalized by the transmission at zero field, $T_{B}/T_{0}$, in which CR manifests itself as an absorption dip.


\begin{figure}[t]
\includegraphics[width=8cm]{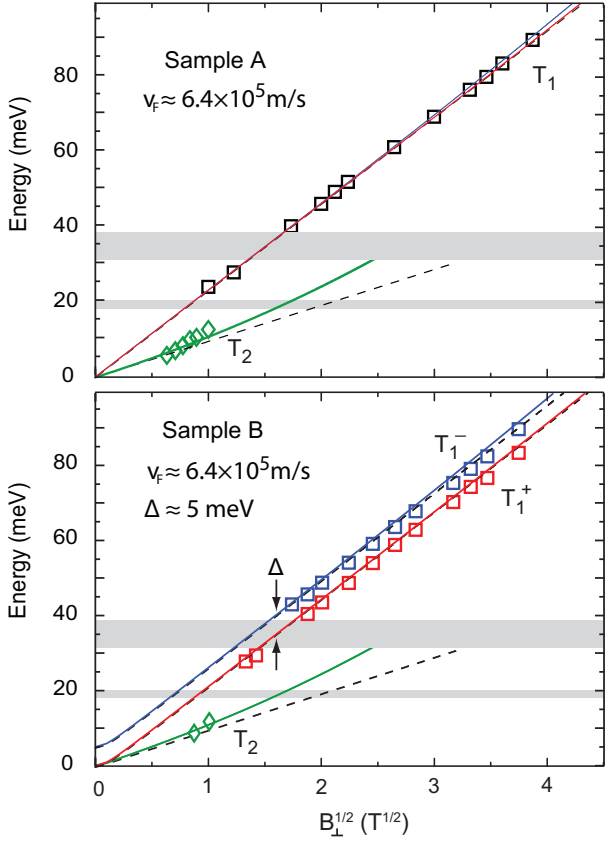}
\caption{
The CR energies plotted as a function of $\sqrt{B_\perp}$ for (a) sample A and (b) sample B. Dashed lines represent the fits using the Dirac LL dispersion (Eq. (\ref{DiracLLmassive})) with (a) $\mathcal{M}=0$ or (b) $\mathcal{M}=2.5$~meV. Solid lines show the transition energies calculated from Eqs. (\ref{En}-\ref{Ezero}) using the parameters described in the text.
}
\label{CRenergies}
\end{figure}

The normalized magneto-transmission spectra of the HgTe QWs are presented in Fig. \ref{FigSpectra}. Outside the opaque reststrahlen-band regions of CdTe (150-160)~cm$^{-1}$ and GaAs (270-310)~cm$^{-1}$, the spectra show the characteristic $\sqrt{B_\perp}$-dependent CR lines (see also Fig. \ref{CRenergies}), expected for 2D Dirac fermions. At low magnetic fields, below 1~T, both samples display very similar behavior, exhibiting a single CR line. In sample A, the CR line is visible down to the lowest measured field of $B_\perp = 0.4$~T, allowing us to estimate a mobility of $\mu>2.5\times 10^{4}$~cm$^{2}$/Vs using the semiclassical condition, $\mu B_{\perp} \geq 1$. We note that, while this value is several times smaller than the mobilities reported in similar QWs \cite{Buttner:2011, Konig:2007, Tkachov:2011, Zholudev:2012, Olbrich:2013, Kvon:2011}, we were unable to measure lower fields because the CR line moves into an opaque region of the spectra.


When the magnetic field increases above $\sim1$~T, the low-field CR series (labeled $T_2$) disappears and a new CR line series ($T_1$) becomes dominant. In Fig. \ref{CRenergies}(a) we plot the CR energies, determined from Lorentzian fits to the spectra of sample A in Fig. \ref{FigSpectra}(a), as a function of $\sqrt{B_\perp}$. Here the $T_1$ CR series exhibits a clear linear dependence in a very broad range of magnetic fields (or equivalently transition energies). The extrapolated linear trend intersects the $y$-axis at zero energy. Contrary to the single CR line behavior in sample A ($d=6.6$~nm), the high-field CR in sample B ($d=6.5$~ nm) splits into two components. These two high-field CR transitions, $T_1^{\pm}$, also show a linear relationship with $\sqrt{B_\perp}$, forming two parallel lines that have the same slope as the $T_1$ line in sample A but shifted by $\pm 2.5$~meV, as shown in Fig. \ref{CRenergies}(b).


Considering the selection rules of IR-active transitions in graphene \cite{Jiang:2007CR,Gusynin:2007}, $| \Delta n |=1$, and n-type doping in our samples, the observed CR series ($T_1$ and $T_2$) can be assigned to the ($0$$\to$$1$) and ($1$$\to$$2$) inter-LL transitions, respectively. At $B>1$~T, Fermi energy shifts from the $n$=2 LL to $n$=1 LL due to the increasing LL degeneracy with the magnetic field, enabling the ($0$$\to$$1$) transition but blocking the ($1$$\to$$2$) transition. The $T_1$ series in both samples perfectly fits the transition energy between $n$=0 and $n$=1 LLs of massive Dirac fermions:
 \begin{equation}
E_n =\mathrm{sgn}(n)\sqrt{\mathcal{M}^2+2|n| v_F^2\hbar^2/l_B^2}.
\label{DiracLLmassive}
 \end{equation}
Good agreement with the data is illustrated in Fig. \ref{CRenergies}(a,b), where the dashed lines plot the CR energies calculated with the Dirac mass $\mathcal{M}$=0 (sample A) or $\mathcal{M}$=2.5~meV (sample B) and  the same band velocity, $v_F=6.4 \times10^{5}$~m/s. The $T_2$ series, however, does not follow the expected $\sqrt{B}$ dependence.


Next we show that all observed CR transitions can be quantitatively described within the analytical model of Bernevig, Hughes and Zhang \cite{Bernevig:2006}. Following \cite{Bernevig:2006,Konig:2007,Konig:2008,Buttner:2011}, the effective Hamiltonian for the lowest energy electron ($E_1$) and hole ($H_1$) subbands near the $\Gamma$-point consists of two uncoupled blocks
 \begin{equation}
\begin{aligned}
&\mathcal{H} = \left( \begin{array}{ccc} H(\mathbf{p}) & 0 \\ 0 & H^*(-\mathbf{p}) \\ \end{array} \right),  \quad  \mathbf{p} = (p_x, p_y, 0) \\
&H(\mathbf{p}) = (\mathcal{C}-\mathcal{D} \mathbf{p} ^2) \mathds{1} + (\mathcal{M}-\mathcal{B} \mathbf{p} ^2)  \sigma_3 + \mathcal{A}(p_x \sigma_x + p_y \sigma_y)
\end{aligned}
\label{EffHamiltonian}
\end{equation}
where $\mathds{1}$ denotes the unit matrix and $\sigma_j$ are the Pauli matrices acting in the $E_1$ and $H_1$ subspace \cite{Bernevig:2006}. The material parameters $\mathcal{A}, \mathcal{B}, \mathcal{C}, \mathcal{D}$, and $\mathcal{M}$ generally depend on the QW thickness and the substrate orientation. The only parameter which changes sign at the critical thickness of the QW is the Dirac mass $\mathcal{M}$. In the presence of an external magnetic field $B_{\perp}$ oriented perpendicular to the QW, as in our experiment, we make the Peierls substitution, ${\bf p}\rightarrow {\bf p}-\frac{e}{c}{\bf A}$, where in the Landau gauge ${\bf A}=B_{\perp}y\hat{x}$. In addition, we add the Zeeman term \cite{Konig:2008,Buttner:2011}, $\frac{1}{2}\mu_BB_{\perp}\left(\begin{array}{cc}g & 0\\ 0 & -g\end{array}\right)$, where $\mu_B$ is the Bohr magneton, $g=\left(\begin{array}{cc}g_e & 0\\ 0 & g_h\end{array}\right)$, and $g_e$($g_h$) are the effective $g$-factor for $E_1$($H_1$) subbands.

Solving the eigenvalue problem \cite{Konig:2008,Buttner:2011} for the upper block gives LL energies for $n \geq 1$:
\begin{eqnarray}
&&E_n^{s}  = \mathcal{C} -(2\mathcal{D}n + \mathcal{B})\frac{ \hbar^2 }{ \ell_B^2}  +s \frac{\mu_{B} B_{\perp}}{4}(g_{e} +g_{h})+ \label{En}\\
&&\alpha \sqrt{2 \mathcal{A}^2 n  \frac{ \hbar^2 }{ \ell_B^2} + \Big( {\mathcal{M} - (2\mathcal{B}n + \mathcal{D}) \frac{ \hbar^2 }{ \ell_B^2}} - s\frac{\mu_{B} B_{\perp}}{4}(g_{e} - g_{h})\Big)^2}
\nonumber\\
&&E_0^{+}  =  \mathcal{C} +\mathcal{M} - (\mathcal{D} + \mathcal{B}) \frac{ \hbar^2 }{ \ell_B^2} + \frac{\mu_{B} B_{\perp}}{4}g_{e}\\
&&E_0^{-}  =  \mathcal{C} -\mathcal{M} - (\mathcal{D} - \mathcal{B}) \frac{ \hbar^2 }{ \ell_B^2} - \frac{\mu_{B} B_{\perp}}{4}g_{h}
\label{Ezero}
\end{eqnarray}
where $s=\pm1$, $E_n^{+}$($E_n^{-}$) stand for spin-up (spin-down) LLs, $\ell_B=\sqrt{\hbar c/e B_{\perp}}$ is the magnetic length, $\alpha = +1$ for conduction band and $-1$ for valence band. In the following, the reference energy is set to zero ($\mathcal{C}$=0) since it does not factor into energy differences probed by CR.

It is instructive to analyze Eqs.~(\ref{En}-\ref{Ezero}) by first setting $\mathcal{M}$=$\mathcal{B}$=$\mathcal{D}$=$g_{e,h}$=0, when one recovers the familiar LL spectrum for  massless Dirac fermions, $E_n =\mathrm{sgn}(n)\sqrt{2 n}\mathcal{A}\hbar/\ell_B$ with the band velocity $v_F=\mathcal{A}$. Lifting the restriction $\mathcal{M}$=0 opens a gap at the $\Gamma$-point of $\Delta E = 2\mathcal{M}$, which can be positive for conventional band structure ($\mathcal{M}$$>$0) or negative for inverted band structure ($\mathcal{M}$$<$0). In the small $B_{\perp}$-field limit, the $\mathcal{A}$($\sqrt{B_\perp}$) term dominates over $\mathcal{B}$ and $\mathcal{D}$ (terms linear in $B_{\perp}$), and the system behaves like a massive Dirac fermion with the Dirac mass $\mathcal{M}$, band velocity $\mathcal{A}$ and an effective $g$-factor reflecting the $B_{\perp}$ linear contribution. When plotted as a function of $\sqrt{B_\perp}$, the energies of the fundamental CR associated with ($0^{\pm}\to1^{\pm}$) transitions appear as two nearly parallel lines and their separation provides a precise measure of the gap and the Dirac mass.
It is worth mentioning that the observed double-peak CR structure can be explained using either $\mathcal{M}=2.5$ or $\mathcal{M}=-2.5$~meV. In other words, our data does not allow one to distinguish between the case when the system has a small normal gap or small inverted gap. However, based on nominal thickness estimations the sample with a smaller QW width (sample B) is expected to have a normal, i.e. non-inverted, band structure.


\begin{figure}[t]
\includegraphics[width=8cm]{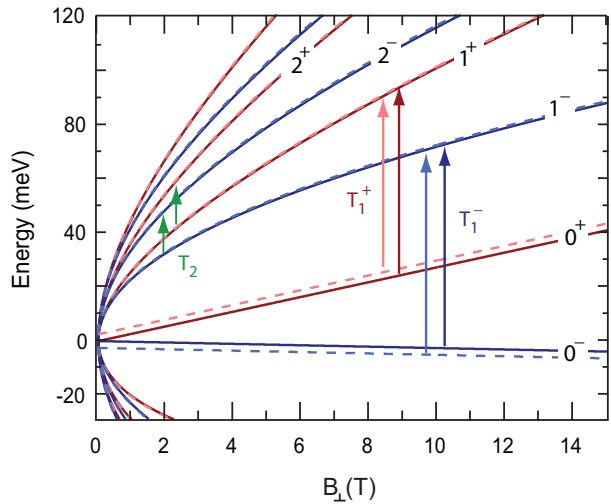}
\caption{
Dispersion of LLs for 2D Dirac fermions calculated from Eqs. (\ref{En}-\ref{Ezero}) for $\mathcal{M}$=0 (solid lines) and $\mathcal{M}$=2.5~meV (dashed lines). Other parameters are described in the text. Arrows indicate the CR transitions observed in this work.
}
\label{LLspectrum}
\end{figure}



Fitting the ($0$$\to$$1$) and ($1$$\to$$2$) transition energies to the experiment, we find that the effective Dirac Hamiltonian of Eq.~(\ref{EffHamiltonian}) describes the data for both samples very well by varying only the Dirac mass $\mathcal{M}$, while keeping all other parameters identical ($\mathcal{A}$=6.4$\times 10^{5}$~m/s, $B$=495~meV nm$^2/\hbar^2$, $D$=345~meV nm$^2/\hbar^2 $, $g_e$=50, and $g_h$=1). We only observe a single CR line in  the $T_2$ series because at low magnetic fields, $B_\perp<1$~T, the spin-up and spin-down ($1$$\to$$2$) transition energies differ by less than $0.1$~cm$^{-1}$, which is much smaller than the CR line width.

In Fig. \ref{LLspectrum} we plot the calculated LL spectrum for both samples. Notice that addition of the mass term in sample B causes the $n=0^{+}(0^{-})$ LL to shift up (down), while the LLs for $n$$>$1 are relatively unaffected. The CR transitions probed in this work are marked by arrows. The corresponding CR energies are plotted as solid lines in Fig. \ref{CRenergies}, demonstrating excellent agreement with the experimental data. Moreover, this model accurately reproduces the field dependence of the CR transitions involving conduction band LLs calculated from the 8$\times$8 Kane model for similar narrow-gap HgTe QWs grown on [013]-oriented GaAs substrates reported in Ref. \cite{Zholudev:2012} by simply adjusting the value of a Dirac mass ($\mathcal{M}$=7~meV). However, this approach fails to represent the LLs in the valence band. This is expected since the model does not take into account the strong conduction/valence band asymmetry. Close to the critical thickness, it fails to describe the valence band LLs at magnetic fields above $\sim2$~T due to a strong deviation from the linear band structure for energies above 30~meV.

In summary, we have performed CR measurements in gapless HgTe QW and observed a characteristic $\sqrt{B}$ dependence of CR energies in high magnetic fields, signifying the presence of massless Dirac fermions in a single valley system. When the width of the QW deviates slightly from the critical thickness, a small band gap occurs along with an effective Dirac mass. We show that the low-energy CR transitions of HgTe QWs near or at the critical thickness can be quantitatively described by an effective Dirac model, making it a valuable tool for future band-gap engineering of similar materials.

\begin{acknowledgments}
This work is supported by the DOE (DE-FG02-07ER46451). Yu.B.V. and N.N.M. acknowledge supports from the Russian Foundation for Basic Research (Grant No. 13-02-00326 a) and the Russian Academy of Sciences.
The CR measurements were performed at the National High Magnetic Field Laboratory, which is supported by the NSF (DMR-0654118), by the State of Florida, and by the DOE. O.V. was supported by the NSF CAREER award under
Grant No. DMR-0955561.
\end{acknowledgments}


\begin{thebibliography}{99}

\bibitem{Novoselov:2005} K. S. Novoselov, A. K. Geim, S. V. Morozov, D. Jiang, M. I. Katsnelson, I. V. Grigorieva, S. V. Dubonos, and A. A. Firsov, Nature \textbf{438}, 197 (2005).

\bibitem{Zhang:2005} Y. Zhang, Y.-W. Tan,  H. L. Stormer, and P. Kim, Nature \textbf{438}, 201 (2005).

\bibitem{Bernevig:2006} B. A. Bernevig, T. L. Hughes, and S. C. Zhang, Science \textbf{314}, 1757 (2006).

\bibitem{Konig:2008} M. K\"onig, H. Buhmann, L. W. Molenkamp, T. Hughes, C.-X. Liu, X. L. Qi, and S. C. Zhang, JPSJ \textbf{77}, 031007 (2008).

\bibitem{Konig:2007} M. K\"onig, S. Wiedmann, C. Br\"une, A. Roth, H. Buhmann, L. W. Molenkamp, X. L. Qi, and S. C. Zhang, Science \textbf{318}, 766 (2007).

\bibitem{Buttner:2011} B. B\"{u}ttner, C. X. Liu, G. Tkachov, E. G. Novik, C. Br\"{u}ne, H. Buhmann, E. M. Hankiewicz, P. Recher, B. Trauzettel, S. C. Zhang, and L. W. Molenkamp, Nature Phys. \textbf{7}, 418 (2011).

\bibitem{Tkachov:2011} G. Tkachov, C. Thienel, V. Pinneker, B. B\"{u}ttner, C. Br\"{u}ne, H. Buhmann, L. W. Molenkamp, and E. M. Hankiewicz, Phys. Rev. Lett. \textbf{106}, 076802 (2011).

\bibitem{McClure:1956} J. W. McClure, Phys. Rev. \textbf{104}, 666 (1956).

\bibitem{Semenoff:1984} G. W. Semenoff, Phys. Rev. Lett. \textbf{53}, 2449 (1984).

\bibitem{Haldane:1988} F. D. M. Haldane, Phys. Rev. Lett. \textbf{61}, 2015 (1988).

\bibitem{Jiang:2007CR} Z. Jiang, E. A. Henriksen, L. C. Tung, Y. J. Wang, M. E. Schwartz, M. Y. Han, P. Kim, and H. L. Stormer, Phys. Rev. Lett. \textbf{98}, 197403 (2007).

\bibitem{Deacon:2007CR} R. S. Deacon, K. C. Chuang, R. J. Nicholas, K. S. Novoselov, and A. K. Geim, Phys. Rev. B \textbf{76}, 081406(R) (2007).

\bibitem{Orlita:2011} M. Orlita, K. Masztalerz, C. Faugeras, M. Potemski, E. G. Novik, C. Br\"une, H. Buhmann, and L. W. Molenkamp, Phys. Rev. B \textbf{83}, 115307 (2011).

\bibitem{Zholudev:2012} M. Zholudev, F. Teppe, M. Orlita, C. Consejo, J. Torres, N. Dyakonova, M. Czapkiewicz, J. Wr\'obel, G. Grabecki, N. Mikhailov, S. Dvoretskii, A. Ikonnikov, K. Spirin, V. Aleshkin, V. Gavrilenko, and W. Knap, Phys. Rev. B \textbf{86}, 205420 (2012).

\bibitem{Ikonnikov:2011} A. V. Ikonnikov, M. S. Zholudev, K. E. Spirin, A. A. Lastovkin, K. V. Maremyanin, V. Ya. Aleshkin, V. I. Gavrilenko, O. Drachenko, M. Helm, J. Wosnitza, M. Goiran, N. N. Mikhailov, S. A. Dvoretskii, F. Teppe, N. Diakonova, C. Consejo, B. Chenaud, and W. Knap, Semicond. Sci. Technol. \textbf{26} 125011 (2011).

\bibitem{Kvon:2011} Z. D. Kvon, S. N. Danilov, D. A. Kozlov, C. Zoth, N. N. Mikhailov, S. A. Dvoretskii, and S. D. Ganichev, JETP Lett. \textbf{94}, 816 (2012).

\bibitem{Olbrich:2013} P. Olbrich, C. Zoth, P. Vierling, K.-M. Dantscher, G. V. Budkin, S. A. Tarasenko, V. V. Bel'kov, D. A. Kozlov, Z. D. Kvon, N. N. Mikhailov, S. A. Dvoretsky, and S. D. Ganichev, Phys. Rev. B \textbf{87}, 235439 (2013).

\bibitem{Gusynin:2007} V. P. Gusynin, S. G. Sharapov, and J. P. Carbotte, Phys. Rev. Lett. \textbf{98}, 157402 (2007).



\end{thebibliography}
\end{document}